\documentclass[twocolumn,english,aps,pre,showpacs]{revtex4}
\usepackage[T1]{fontenc}
\usepackage[latin1]{inputenc}
\usepackage{amsmath}
\usepackage{graphicx}
\usepackage{amssymb}

\makeatletter
\usepackage{hyperref}
\hypersetup{backref, 
colorlinks=true,
citecolor=blue,
linkcolor=blue,
filecolor=blue,
urlcolor=blue
}

\usepackage{babel}
\makeatother
\begin{document}

\title{Weakly disordered absorbing-state phase transitions}

\author{Jos\'e A. Hoyos}

\affiliation{Department of Physics, Missouri University of Science and Technology,
Rolla, Missouri 65409, USA}

\affiliation{Department of Physics, Duke University, Durham, North Carolina 27708,
USA}

\begin{abstract}
The effects of quenched disorder on nonequilibrium phase transitions
in the directed percolation universality class are revisited. Using
a strong-disorder energy-space renormalization group method, it is
shown that for any amount of disorder the critical behavior is controlled
by an infinite-randomness fixed point in the same universality class
of the random transverse-field Ising models.
\end{abstract}


\maketitle
Using the formalism and the knowledge of equilibrium phase transitions,
a natural trend with the aim of establishing and classifying possible
universality classes in nonequilibrium transitions arose \cite{marro-dickman-book,dickman-97}.
It was conjectured that the critical behavior of short-ranged interacting
models with scalar order parameter and absence of conservation laws
and extra symmetries are in the Directed Percolation universality
class \cite{grassberger-delatorre-ap79,janssen-zpb81,grassberger-zpb82},
which separates an active fluctuating state from an inactive (absorbing)
nonfluctuating one \cite{hinrichsen-review00}. Examples include transitions
in the contact process \cite{harris-an74}, catalytic reactions \cite{ziff-gulari-barshad-prl86},
depinning interface growth \cite{tang-leschhorn-pra92,buldyrev-etal-pra92},
and marginal growth of turbulent domains in laminar flows \cite{pomeau-pd86}.

Despite the theoretical understanding on the ubiquitous Directed Percolation
universality class, its critical exponents have hardly been seen in
real experiments \cite{hinrichsen-bjp00} (see, however, Ref.~\cite{takeuchi-etal-prl07}).
It was then suspected that quenched disorder may be responsible. For
spatial dimension $d<4$, this is indeed the case as dictated by the
Harris criterion \cite{harris-jpc74,kinzel-zpb85,noest-prl86} and
confirmed by field-theoretical methods \cite{janssen-pre97}, which
showed that the renormalization-group equations have only runaway
solutions towards large disorder. In addition, disorder-dependent
Griffiths-like phases \cite{griffiths-prl69,mccoy-prl69} nearby criticality
have been observed \cite{noest-prl86,noest-prb88,moreira-dickman-pre96,dickman-moreira-pre98,bramson-durrett-shonmann-ap90,webman-etal-pmb98,cafiero-gabrielli-munoz-pre98}. 

This scenario thus points out an unconventional critical behavior
originating from the interplay between \emph{large} spatial disorder
fluctuations and strong correlations. Motivated by this reasoning,
a strong-disorder renormalization-group (SDRG) method \cite{MDH-PRL,MDH-PRB}
was applied to the random contact process model \cite{hooyberghs-prl}.
For \emph{strong} disorder, the critical behavior is governed by a
universal infinite-randomness fixed point (IRFP) in the same universality
class of the random transverse-field Ising model, whose dynamical
scaling is known to be activated \cite{fisher92,fisher95,motrunich-ising2d},
i.e., length $\xi$ and time $\tau$ are related through $\ln\tau\sim\xi^{\psi}$,
with $\psi$ (dubbed tunneling exponent) being universal. For \emph{weak}
disorder, on the other hand, the critical point has finite disorder
and usual power-law scaling $\tau\sim\xi^{z}$ with nonuniversal dynamical
exponent $z$ proportional to the disorder strength and is formally
infinite at the transition between the weak- and strong-disorder limits.
These conclusions were also supported by density-matrix renormalization-group
calculations in $d=1$ \cite{hooyberghs-prl}. Further Monte-Carlo
calculations in $d=2$ confirmed the above scenario. However, the
possibility that the weak-disorder regime was an artifact of finite-size
effects was raised \cite{hooyberghs-pre}.

Facing the logarithmically slow dynamics, large-scale Monte-Carlo
simulations in $d=1$ for system sizes up to $10^{7}$ sites and times
up to $10^{9}$ were performed \cite{vojta-dickison-nonequ}. No nonuniversal
weak-disorder critical regime was found, shrinking considerably the
parameter space in which it would exist and, together with the field-theoretical
results, strongly suggesting its nonexistence. It then raises the
following puzzle. How can the SDRG suggest a \emph{finite}-disordered
fixed point while Monte-Carlo simulations point to an \emph{infinite}-disordered
one? Since the SDRG method is devised to include any minimal effects
of disorder, it should be able to capture the physics of any IRFP
as well as to point out its existence.

This Brief Report is devoted to solve this question. In generalizing
the SDRG method, we show that the critical system is governed by a
universal IRFP when \emph{any} amount of disorder is present. Moreover,
our motivation goes beyond the issue of settling the correct universality
class of weakly disordered absorbing-state phase transitions. It deals
with the delicate issue of implementing a SDRG in such a limit, which
is an important tool to tackle many disordered systems.

For definiteness, we now introduce the system, review the usual SDRG
for random contact process \cite{hooyberghs-prl}, point out its failure,
and modify it in order to overcome this problem.

The contact process can be defined in a lattice in which each site
$i$ can have either a healed $(\sigma_{i}=1)$ or an infected $(\sigma_{i}=-1)$
particle. A healed particle at site $i$ can be contaminated by an
infected one in a neighboring site $j$ at rate $\lambda_{ij}=\lambda_{ji}$
\cite{foot1}. Also, an infected particle at site $i$ can get spontaneously healed
at rate $\mu_{i}$. The system has a stochastic dynamics governed
by a master equation $\partial_{t}\mathbf{P}(\{\sigma\},t)=-H\mathbf{P}(\{\sigma\},t)$
where the vector $\mathbf{P}$ gives the probability of finding the
configuration $\{\sigma\}=(\sigma_{1},\sigma_{2},\dots)$ at time
$t$ and \begin{equation}
H=\sum_{i}\mu_{i}M_{i}+\sum_{\left\langle i,j\right\rangle }\lambda_{ij}\left(n_{i}Q_{j}+Q_{i}n_{j}\right)\label{eq:H}\end{equation}
 is the generator of the Markov process \cite{alcaraz-etal-ap94,schutz-inPTCP,hooyberghs-pre}.
Here, \[
M=\left(\begin{array}{cc}
0 & -1\\
0 & 1\end{array}\right)~,~~n=\left(\begin{array}{cc}
0 & 0\\
0 & 1\end{array}\right)~,~~Q=\left(\begin{array}{cc}
1 & 0\\
-1 & 0\end{array}\right)~,\]
and $\left\langle i,j\right\rangle $ restricts the sum to nearest
neighbors only.

The usefulness of this ``quantum Hamiltonian formalism'' comes from
the fact that the steady state probability distribution $\mathbf{P}(\{\sigma\},t\rightarrow\infty)$
coincides with ground state of $H$ and that the long-time relaxation
properties are obtained from the low-lying spectrum of $H$. Although
$H$ is in general non-Hermitian, some standard methods can still
be used.

For the disordered case, $\lambda_{ij}$ and $\mu_{i}$ are random
independent variables distributed according to $P(\lambda)$ and $R(\mu)$,
respectively. In this case, the low-lying spectrum of $H$ can be
reached by the following recipe (for simplicity, we focus on the $d=1$
case): (i) Search for the fastest (``high-energy'') scale in the system
$\Omega=\max\{\lambda_{i},\mu_{i}\}$, (ii) integrate out locally
the corresponding mode, and (iii) renormalize the remaining degrees
of freedom. Those steps are the basis of the SDRG method \cite{igloi-review}.

When (ii.a) $\Omega=\lambda_{2}$, particles on sites 2 and 3 can
be considered as one since they will be mostly in the same state,
i.e., either both healed or both infected. Then, (iii.a) one treats
$H_{0}=\lambda_{2}(n_{2}Q_{3}+Q_{2}n_{3})$ exactly and $H_{1}=\mu_{2}M_{2}+\mu_{3}M_{3}$
as a perturbation. $H_{0}$ has two twofold multiplets. In the ground
(excited) one, particles 2 and 3 are in the same (opposite) state.
$H_{1}$ lifts the degeneracy of the ground multiplet, which corresponds
to the effective healing rate $\tilde{\mu}$ of the particle cluster
2 and 3. In second order of perturbation theory, one finds that $\tilde{H}_{1}=\tilde{\mu}\tilde{M}$,
with\begin{equation}
\tilde{\mu}=\kappa_{\mu}\mu_{2}\mu_{3}/\lambda_{2}~,\textrm{ with }\kappa_{\mu}=2~.\label{eq:mu-tilde}\end{equation}

When (ii.b) $\Omega=\mu_{2}$, the particle at site 2 can be considered
as healed for all times. Hence, (iii.b) one treats $H_{0}=\mu_{2}M_{2}$
exactly and $H_{1}=\lambda_{1}(n_{1}Q_{2}+Q_{1}n_{2})+\lambda_{2}(n_{2}Q_{3}+Q_{2}n_{3})$
perturbatively. $H_{0}$ has two fourfold multiplets. The ground (excited)
one refers to particle 2 healed (infected). $H_{1}$ then lifts the
degeneracy which corresponds to an effective infection rate $\tilde{\lambda}$
between particles 1 and 3. In second order of perturbation theory
$\tilde{H}_{1}=\tilde{\lambda}(n_{1}Q_{3}+Q_{1}n_{3})$, with\begin{equation}
\tilde{\lambda}=\kappa_{\lambda}\lambda_{1}\lambda_{2}/\mu_{2}~,\textrm{ with }\kappa_{\lambda}=1~.\label{eq:lambda-tilde}\end{equation}

Once set the recursion relations (\ref{eq:mu-tilde}) and (\ref{eq:lambda-tilde}),
flow equations for $P(\lambda)$ and $R(\mu)$ can be constructed
and the fixed-point distributions obtained \cite{hooyberghs-pre,fisher95}.
In principle, this give the long-time behavior of the system. The
multiplicative structure of Eqs.~(\ref{eq:mu-tilde}) and (\ref{eq:lambda-tilde})
is very important. Under these transformations, $P(\lambda)$ and
$R(\mu)$ become indefinitely broad at criticality for any amount
of disorder as long as $0<\kappa_{\mu,\lambda}\leq1$ \cite{fisher94-xxz,fisher95,hooyberghs-pre}.
However, for $\kappa_{\mu,\lambda}>1$ the SDRG becomes inconsistent
for weak disorder because the renormalized couplings are typically
bigger than the decimated ones. It is thus tempting to interpret this
result as a runaway flow towards weak disorder in odds with the field-theoretical
results \cite{janssen-pre97}. As we show below, this is not the case.
The generation of a transition rate larger than the decimated ones
is unphysical. The numerical prefactor $\kappa_{\mu}>1$ is just an
artifact of treating $H_{1}$ until second order in perturbation theory.

According to Eq.~(\ref{eq:mu-tilde}), the splitting of the ground
multiplet of $H_{0}$ due to $H_{1}$ may overcome the distance ($\lambda_{2}$)
between the two unperturbed multiplets for certain values of $\mu_{2,3}$
even though $\mu_{2,3}<\lambda_{2}$. Treating $H_{0}+H_{1}$ exactly,
however, this can never be the case. The ground state has energy $0$
and the excited ones are solutions of the polynomial \begin{equation}
x^{3}-2\xi x^{2}+\left(\xi^{2}+\mu_{2}\mu_{3}\right)x-\mu_{2}\mu_{3}\left(\lambda_{2}+\xi\right)=0~,\label{eq:poly}\end{equation}
with $\xi=\lambda_{2}+\mu_{2}+\mu_{3}$. The renormalized healing
rate $\tilde{\mu}$ is thus its minimal root. Although we could not
solve $\tilde{\mu}$ analytically, its maximum value is shown to be
$\tilde{\mu}_{{\rm max}}=(2-\sqrt{2})\lambda_{2}$, which happens
for $\mu_{3}=\mu_{2}=\lambda_{2}$. Numerical inspections of Eq.~(\ref{eq:poly})
show that $\tilde{\mu}\leq\min\{\mu_{2},\mu_{3},\lambda_{2}\}$ in
general. 

\begin{figure}[t]
\begin{center}\includegraphics[%
  clip,
  width=1\columnwidth,
  keepaspectratio]{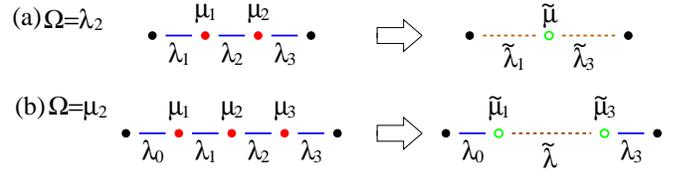}\end{center}

\caption{(Color online) Schematic decimation procedure.\label{cap:Schematic-decimation-procedure.}}
\end{figure}

In addition, the operators connecting this particle cluster to the
rest of the chain have also to be projected onto the same states.
We find that $n_{2,3}=\alpha_{2,3}\tilde{n}$ and $Q_{2,3}=\alpha_{2,3}\tilde{Q}$,
where $(1+c_{2}+c_{3})\alpha_{2,3}=1+c_{2,3}$, with $(\lambda_{2}+\mu_{2,3}-\tilde{\mu})^{2}c_{2,3}=\lambda_{2}\mu_{3,2}$.
(Note that $1/2\leq\alpha_{2,3}\leq1$.) Therefore, the SDRG decimation
procedure summarizes in replacing $\sum_{i=1,3}\lambda_{i}(n_{i}Q_{i+1}+Q_{i}n_{i+1})+\sum_{i=2,3}\mu_{i}M_{i}$
by $\tilde{\lambda}_{1}(n_{1}\tilde{Q}+Q_{1}\tilde{n})+\tilde{\mu}\tilde{Q}+\tilde{\lambda}_{3}(\tilde{n}Q_{4}+\tilde{Q}n_{4}),$
with $\tilde{\lambda}_{1,3}=\alpha_{2,3}\lambda_{1,3}$ {[}see Fig.~\hyperref[cap:Schematic-decimation-procedure.]{\ref{cap:Schematic-decimation-procedure.}(a)}{]}.
The renormalization of $\lambda_{1,3}$ is not considered in the usual
perturbative SDRG which is indeed a ``weaker'' effect since $\alpha_{2,3}\in[1/2,1]$
and approaches $1$ in the strong-disorder limit. 

Repeating the same procedure when decimating a healing rate, Eq.~(\ref{eq:lambda-tilde})
is then replaced by \begin{equation}
\tilde{\lambda}=\zeta-\chi~,\label{eq:lambda-tilde-exact}\end{equation}
with $2\zeta=\lambda_{1}+\lambda_{2}+\mu_{2}$ and $\chi=\sqrt{\zeta^{2}-\lambda_{1}\lambda_{2}}$,
implying $\tilde{\lambda}\leq\min\{\lambda_{1},\lambda_{2},\mu_{2}\}$.
{[}Its maximal value $\tilde{\lambda}_{{\rm max}}=\mu_{2}(3-\sqrt{5})/2$
happens for $\lambda_{1}=\lambda_{2}=\mu_{2}$.{]} Moreover, $M_{1,3}=\beta_{1,3}\tilde{M}_{1,3}$,
$n_{1,3}=\tilde{n}_{1,3}$ and $Q_{1,3}=\tilde{Q}_{1,3}$, with $4\beta_{1}\chi(\zeta+\chi)=\lambda_{1}(3\mu_{2}-\lambda_{2})+(\lambda_{2}+\mu_{2})(\mu_{2}+\lambda_{2}+2\chi)$
and $\beta_{3}$ is obtained by exchanging $\lambda_{1}\rightleftharpoons\lambda_{2}$
in $\beta_{1}$. (Note that $3/4\leq\beta_{1,3}\leq1$.) These results
mean we have to replace $\sum_{i=0,3}\lambda_{i}(n_{i}Q_{i+1}+Q_{i}n_{i+1})+\sum_{i=1,3}\mu_{i}M_{i}$
by $\lambda_{0}(n_{0}\tilde{Q}_{1}+Q_{0}\tilde{n}_{1})+\tilde{\mu}_{1}\tilde{M}_{1}+\tilde{\lambda}(\tilde{n}_{1}\tilde{Q}_{2}+\tilde{Q}_{1}\tilde{n}_{2})+\tilde{\mu}_{3}\tilde{M}_{3}+\lambda_{3}(\tilde{n}_{3}Q_{4}+\tilde{Q}_{3}n_{4})$,
where $\tilde{\mu}_{1,3}=\beta_{1,3}\mu_{1,3}$ {[}see Fig.~\hyperref[cap:Schematic-decimation-procedure.]{\ref{cap:Schematic-decimation-procedure.}(b)}{]}.
Note that $n_{1,3}$ ($Q_{1,3}$) has no projection onto $\tilde{n}_{3,1}$
($\tilde{Q}_{3,1}$). If this was not the case, the technical treatment
of this SDRG would be more difficult because further-nearest-neighbor
interactions would arise. Long-ranged interactions may point out delocalized
states. Their absence suggests that the SDRG here presented is amenable.

Importantly, there are no level crossings in the entire region where
the parameters of $H_{1}$ are less than or equal to the parameters
of $H_{0}$, meaning the interpretation of the decimation steps still
holds. Also important, the energy difference between the second and
first excited multiplets ($\Delta_{21}$) of $H_{0}+H_{1}$ only increases
when increasing the perturbation and is always greater than the energy
difference between the first excited and ground multiplets ($\Delta_{10}$).
Precisely, $\Delta_{10}\leq\left(1-1/\sqrt{2}\right)\Delta_{21}$.

Therefore, exactly projecting the entire system in the low-energy
states of $H_{0}+H_{1}$ makes the renormalization-group approach
totally consistent. Whether or not these new recursion relations drive
the system to the universal IRFP is not straightforwardly clear. This
is the question we address in the next part of this paper.

The fate of the critical point is obtained by solving the standard
flow equations \cite{fisher95,hooyberghs-pre} for $P(\lambda)$ and
$R(\mu)$ with the perturbed renormalized rates (\ref{eq:mu-tilde})
and (\ref{eq:lambda-tilde}) replaced by their exact counterparts
(\ref{eq:poly}) and (\ref{eq:lambda-tilde-exact}) in addition to
the weaker renormalization of the neighboring transition rates ($\tilde{\lambda}_{1,3}$
and $\tilde{\mu}_{1,3}$ in Fig.~\ref{cap:Schematic-decimation-procedure.}).
Because of the complicated analytical structure of these quantities,
a detailed analytical solution is hampered. We then rewrite $\tilde{\mu}=\kappa_{\mu}^{\prime}\mu_{2}\mu_{3}/\lambda_{2}$
and $\tilde{\lambda}=\kappa_{\lambda}^{\prime}\lambda_{1}\lambda_{2}/\mu_{2}$,
where $\kappa_{\mu,\lambda}^{\prime}$ are functions of the decimated
transition rates. Moreover, we will neglect the renormalizations of
$\tilde{\lambda}_{1,3}$ and $\tilde{\mu}_{1,3}$ %
\cite{foot2}. Now, recall that (i) $\tilde{\lambda}$ and $\tilde{\mu}$ are always
less than the decimated ones and that (ii) there is no correction
to $H_{0}$ in first order of perturbation theory. Point (i) permits
us to set $\kappa_{\mu,\lambda}^{\prime}=1$ in the weak-disorder
limit. Hence, the system rapidly flows towards stronger disorder.
As intermediate disorder is reached, the only way of stopping its
further growth is making \emph{all} decimations of type $\tilde{\mu}={\rm const}\times\mu_{2}$
\cite{hoyos-sun}, which corresponds to corrections in first order
of perturbation theory. Point (ii) thus guarantees there is no hindrance
on the flow towards even stronger disorder, in which limit $\kappa_{\mu,\lambda}^{\prime}$
can be neglected \cite{fisher94-xxz}. We thus finally conclude that
any amount of disorder drives the critical system towards the universal
infinite-randomness fixed point.

\begin{figure}[b]
\begin{center}\includegraphics[%
  clip,
  width=1\columnwidth,
  keepaspectratio]{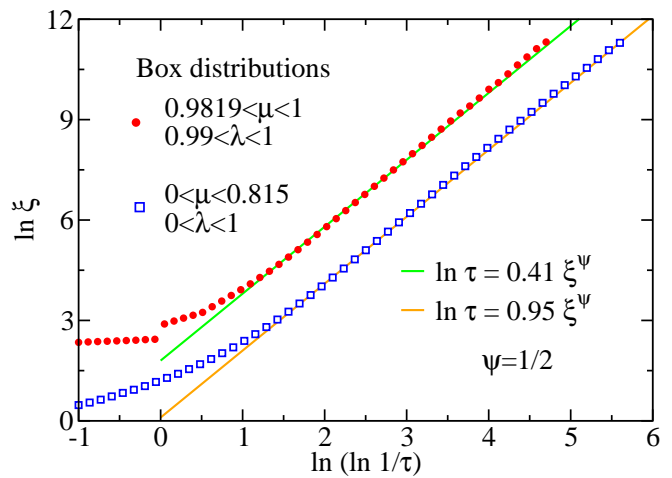}\end{center}

\caption{\label{cap:scaling}(Color online) Time $\tau$ and length $\xi$
scales along the SDRG flow at criticality. Transition rates are drawn
from boxlike distributions as indicated. Chains have $2\times10^{6}$
sites and the data were averaged over $100$ disorder realizations.
Error bars are about the symbol size.}
\end{figure}

This conclusion was checked by numerical implementation of the SDRG
for weak- and moderate-disordered chains. (For consistency with the
above proof, the weaker corrections to $\tilde{\lambda}_{1,3}$ and
$\tilde{\mu}_{1,3}$ were neglected %
\cite{foot3}.) Following time $\tau$ and length $\xi$ scales along the critical
SDRG flow, the predicted \cite{fisher92} tunneling exponent $\psi=1/2$
was confirmed (see Fig.~\ref{cap:scaling}). Here, $\tau^{-1}=\Omega$
and $\xi^{-1}$ is the density of active particle clusters. The off-critical
Griffiths phases surrounding the critical point in which $\tau\sim\xi^{z}$
with disorder-dependent dynamical exponent is also confirmed in our
numerics \cite{foot4}.

We now address the issue of weak disorder in higher dimensions. One
key feature hinders a straightforward generalization of the RG steps
here proposed: the reconnection of the lattice. Because the coordination
number increases in $d>1$, one eventually needs to treat exactly
big particle clusters. Leaving this task open, we cannot guarantee
that all the RG steps will consistently lower the energy scale and
drive the system to an IRFP. However, and somewhat surprisingly, it
was shown that the lattice reconnection does not hinder the flow towards
infinite randomness \cite{motrunich-ising2d}. As the inconsistency
of the simple recursion relations is just an artifact of the perturbative
treatment, it is then reasonable to conclude that the RG flows of
the Directed Percolation and the Transverse-Field Ising universality
classes are the same in the presence of disorder in $d=2$ and $3$,
as suspected in Ref.~\cite{hooyberghs-pre}. This would be in agreement
with Monte-Carlo simulations in $d=2$ \cite{vojta-farquhar-mast},
with the Harris criterion \cite{harris-jpc74,kinzel-zpb85,noest-prl86}
and with the field-theoretical runaway flow solutions \cite{janssen-pre97}.

Recently, the experimental realization of the clean Directed Percolation
universality class have been claimed \cite{rupp-richter-rehberg-pre03,takeuchi-etal-prl07}.
These experiments now raise another puzzle in the face of our results
and many others \cite{harris-jpc74,kinzel-zpb85,noest-prl86,janssen-pre97,hooyberghs-pre,vojta-dickison-nonequ,vojta-farquhar-mast}.
We would like to point out two crossovers which may give an explanation.
One is the time crossover which was stressed in Ref.~\cite{vojta-dickison-nonequ}
(see, \emph{e.g}., Fig.~7 therein). Because of the logarithmically
slow dynamics, the ``true'' steady state takes place only after a
long period of relaxation. The other one is the clean-dirty crossover
length. As in spin chains \cite{laflorencie-correlacao-PRL,laflorencie-correlacao-PRB},
there is a crossover length below which disorder is irrelevant. The
cleaner the sample the longer the crossover length which reaches hundreds
of sites even for spin chains with moderate disorder. The time crossover
is analogous to the temperature crossover in spin chains. Only at
very low temperatures are the low-energy states important. The length
crossover is equally analogous. Statistically rare fluctuations (the
so-called large rare regions) only exist on large samples. Naturally,
these crossovers are related through the dynamics. In Ref.~\cite{rupp-richter-rehberg-pre03},
the system size is of order of hundreds of degrees of freedom. It
is thus reasonable that the exponents measured are nonuniversal between
the clean and the infinite-randomness fixed point. (This also may
apply to other experiments \cite{hinrichsen-bjp00}.) The crossover
length of the samples in Ref.~\cite{takeuchi-etal-prl07} seems much
bigger.

In the face of the possibility of explaining many experiments, it
is thus desirable to study the aforementioned crossover of the exponents,
which should be accomplished without much effort by Monte-Carlo calculations
in $d=1$, for instance. From the experimental side, it is desirable
to pinpoint precisely the source of quenched \emph{}disorder and to
estimate its strength. Border effects may also diminish the effective
size of the sample. Finally, due to the slow relaxation processes,
time measurements have to be carefully taken when locating the critical
point. These studies should shed considerable light on this problem.

In conclusion, we have modified the usual strong-disorder renormalization-group
method in order to \emph{exactly} recast the low-energy spectrum of
the local fast-mode Hamiltonian. This allowed the method amenable
to attack the problem in the weak-disorder limit in which the perturbative
treatment yielded to runaway flow towards weak disorder. Applications
to quantum spin chains as well as comparison with similar generalizations
will be presented elsewhere.

As discussed in Refs.~\cite{fisher94-xxz,fisher95}, this renormalization
group method is not justified in the weak-disorder limit. We, however,
leave open the possibility that, by exactly projecting the entire
Hamiltonian onto the local low-energy spectrum, the method will correctly
point out whether weak disorder is irrelevant.

We are indebit to T. Vojta, E. Miranda, M.-Y. Lee, and K. A. Takeuchi
for useful discussions. This work was supported by the NSF under Grants
Nos. DMR-0339147 and DMR-0506953, and by Research Corporation.

\bibliographystyle{apsrev}
\bibliography{/home/hoyos/Documents/referencias/referencias}

\begin{thebibliography}{43}
\bibitem{foot1} Note we are not dividing $\lambda_{ij}$ by the coordination number as usual.
\bibitem{foot2}If the system flows towards strong disorder (as is the case), considering the renormalization of $\tilde{\lambda}_{1,3}$ and $\tilde{\mu}_{1,3}$ will only support this result.
\bibitem{foot3} Their only effect is to shift the value of the critical point closer to its true value in addition to enhance the crossover length for
weaker disorder. We remember that those nonuniversal quantities are not suitably computed by RG methods.
\end{thebibliography}

\end{document}